\documentclass[twocolumn]{aastex631}
\usepackage{xcolor}
\newcommand\gildas{\textsc{gildas}}
\newcommand\ci{[C\,\textsc{i}]}
\newcommand\cii{[C\,\textsc{ii}]}

\def\arcmin{\ifmmode {^{\prime}}\else $^{\prime}$\fi}
\def\arcsec{\ifmmode {^{\prime\prime} }\else $^{\prime\prime}$\fi}
\usepackage{amsmath,amsthm,amssymb,amsfonts,esint,graphicx,mwe,wrapfig,bm,listings,siunitx,float,hyperref}

\usepackage{multirow}

\received{2024 Dec 23}
\revised{2025 Feb 7}
\accepted{2025 Feb 19}
\begin{document}
\title{A Large Molecular Gas Reservoir in the Protocluster SPT2349$-$56 at $z\,{=}\,4.3$}
\correspondingauthor{Dazhi Zhou}
\email{dzhou.astro@gmail.com}
\author[0000-0002-6922-469X]{Dazhi Zhou}
\affiliation{\ubc}
\author[0000-0002-8487-3153]{Scott~C. Chapman}
\affiliation{Department of Physics and Atmospheric Science, Dalhousie University, Canada}
\affiliation{\ubc}
\affiliation{National Research Council, Herzberg Astronomy and Astrophysics, Canada}
\author[0000-0002-3187-1648]{Nikolaus Sulzenauer}
\affiliation{\mpifr}
\author[0009-0008-8718-0644]{Ryley Hill}
\affiliation{\ubc}
\author[0000-0002-6290-3198]{Manuel Aravena}
\affiliation{\udp}
\author[0000-0003-2860-5717]{Pablo Araya-Araya}
\affiliation{Departamento de Astronomia, Instituto de Astronomia, Geofísica e Ciências Atmosféricas, Universidade de São Paulo, \\Rua do Matão 1226, Cidade Universitária, 05508-900, São Paulo, SP, Brazil}
\author[0000-0002-4657-7679]{Jared Cathey}
\affiliation{\ufastro}
\author[0000-0002-2367-1080]{Daniel~P. Marrone}
\affiliation{Steward Observatory, University of Arizona, 933 North Cherry Avenue, Tucson, AZ 85721, USA}
\author[0000-0001-7946-557X]{Kedar~A. Phadke}
\affiliation{\illinoisastro}
\affiliation{\illinoiscaps}
\author[0000-0001-7477-1586]{Cassie Reuter}
\affiliation{\berkeley}
\author[0000-0001-6629-0379]{Manuel Solimano}
\affiliation{\udp}
\author[0000-0003-3256-5615]{Justin~S.~Spilker}
\affiliation{\tamu} 
\author[0000-0001-7192-3871]{Joaquin~D. Vieira}
\affiliation{\illinoisastro}
\affiliation{\illinoiscaps}
\affiliation{\illinoisphysics}
\author[0000-0001-7610-5544]{David Vizgan}
\affiliation{\illinoisastro}
\author[0009-0003-4626-9777]{George~C.~P. Wang}
\affiliation{\ubc}
\author[0000-0003-4678-3939]{Axel Weiss}
\affiliation{\mpifr}
\newcommand{\illinoisastro}{\text{Department of Astronomy, University of Illinois, 1002 West Green St., Urbana, IL 61801, USA}}
\newcommand{\illinoiscaps}{\text{Center for AstroPhysical Surveys, National Center for Supercomputing Applications, 1205 West Clark Street,}\linebreak \text{Urbana, IL 61801, USA}}
\newcommand{\illinoisphysics}{\text{Department of Physics, University of Illinois, 1110 West Green St., Urbana, IL 61801, USA}}
\newcommand{\ufastro}{\text{Department of Astronomy, University of Florida, Gainesville, FL 32611, USA}}
\newcommand{\tamu}{\text{Department of Physics and Astronomy and George P. and Cynthia Woods Mitchell Institute for Fundamental Physics and Astronomy, }\linebreak \text{Texas A\&M University, 4242 TAMU, College Station, TX 77843-4242, USA}}
\newcommand{\cfa}{\text{Center for Astrophysics | Harvard \& Smithsonian, 60 Garden St, Cambridge, MA 02138, USA}}
\newcommand{\ubc}{\text{Department of Physics and Astronomy, University of British Columbia, 6225 Agricultural Rd., Vancouver, V6T 1Z1, Canada}}
\newcommand{\mpifr}{\text{Max-Planck-Institut für Radioastronomie, Auf dem Hügel 69, 53121 Bonn, Germany}}
\newcommand{\berkeley}{\text{Department of Physics, University of California, 366 Physics North MC 7300, Berkeley, CA, 94720-7300, USA}}
\newcommand{\udp}{\text{Instituto de Estudios Astrof\'isicos, Facultad de Ingenier\'ia y Ciencias, Universidad Diego Portales, }\linebreak \text{Av. Ej\'ercito Libertador 441, Santiago, Chile C\'odigo Postal 8370191}}
\begin{abstract}
We present Atacama Compact Array (ACA) Band-3 observations of the protocluster SPT2349$-$56, an extreme system hosting ${>}\,10$ ultraluminous infrared galaxies (ULIRGs, $L_{\rm IR}\,{\gtrsim}\,10^{12} L_\odot$) in a 200-kpc diameter region at $z\,{=}\,4.3$, to study its integrated molecular gas content via CO(4--3) and long-wavelength dust continuum. 
The $\sim$\,30-hour integration represents one of the longest exposures yet taken on a single pointing with the ACA 7-m. 
The low-resolution ACA data ($21.0\arcsec\,{\times}\,12.2\arcsec$) reveal a 75\% excess CO(4--3) flux compared to the sum of individual sources detected in higher-resolution Atacama Large Millimeter Array (ALMA) data ($1.0\arcsec\,{\times}\,0.8\arcsec$). 
Our work also reveals a similar result by tapering the ALMA data to $10\arcsec$. 
In contrast, the 3.2\,mm dust continuum shows little discrepancy between ACA and ALMA. 
A single-dish {\cii} spectrum obtained by APEX/FLASH supports the ACA CO(4--3) result, revealing a large excess in {\cii} emission relative to ALMA. 
The missing flux is unlikely due to undetected faint sources but instead suggests that high-resolution ALMA observations might miss extended and low-surface-brightness gas. 
Such emission could originate from the circum-galactic medium (CGM) or the pre-heated proto-intracluster medium (proto-ICM). 
If this molecular gas reservoir replenishes the star formation fuel, the overall depletion timescale will exceed 400\,Myr, reducing the requirement for the simultaneous ULIRG activity in SPT2349$-$56. 
Our results highlight the role of an extended gas reservoir in sustaining a high star formation rate (SFR) in SPT2349$-$56, and potentially establishing the ICM during the transition phase to a mature cluster.
\end{abstract}
\section{Introduction} \label{sec:intro}
Galaxy clusters are the largest and the densest structures in the local universe. 
They are characterized by their high galaxy overdensities but low star formation rates (SFRs), where red and quiescent galaxies are expected to be the main population \citep[e.g.,][]{Peebles1980,Dressler1980,Peng2010}. As the high-redshift progenitors of galaxy clusters, protoclusters reveal the epoch previous their virialization when their member galaxies are actively forming stars, thereby providing excellent laboratories for studying dark matter halo assembly and the impact of dense environments on galaxy evolution in the early Universe \citep{Chiang2013,Chiang2017,Overzier2016, Wang2016,wang2024100}. 
Of particular importance are the ultraluminous infrared galaxies (ULIRGs, $L_{\rm IR}\,{\gtrsim}\,10^{12} L_\odot$, $\rm SFR\,{\gtrsim}\,100\,M_\odot/yr$) or submillimeter galaxies (SMGs, $S_{850}\,{\geq}\,1\,\rm mJy$), which represent the most active phase of star-formation in massive galaxies. They are important sites of stellar mass build-up in protoclusters, and are thought to be the precursors of giant elliptical galaxies in galaxy clusters \citep[e.g.,][]{Blain04,Capak2011,Toft2014,casey14,Casey2016,hodge20,Araya-Araya2024}. 

Discovered by the South Pole Telescope (SPT) in the 2500\,deg$^2$ multi-band SPT-SZ millimeter survey \citep{Vieira2010,Carlstrom2011,Mocanu2013}, SPT2349$-$56 ($z\,{=}\,4.3$) is the most intensely star-forming protocluster core at $z\,{\sim}\,4$ confirmed to date \citep{Miller2018, Wang2021}. 
More than 20 protocluster members have been spectroscopically confirmed within a 200-kpc diameter region by submillimeter observations with ALMA, of which at least 11 can be further classified as ULIRGs on the main sequence within the southern core, meaning that this system contains an extremely high overdensity of ULIRGs compared to other known protoclusters \citep{Hill2020,Hill2022,Rotermund2021,Apostolovski2024, Venkateshwaran2024, Hughes2024}. 
In the central core region, multiple active galactic nuclei (AGNs) have been detected and the SFR density exceeds 10,000\,$\rm M_\odot\,yr^{-1}\,Mpc^{-3}$ \citep{Chapman2024,Vito2024}. 
Simulations suggest that most of the central galaxies will merge to form the brightest cluster galaxy (BCG) within 1 billion years \citep{Rennehan2020}.

The extreme nature of such protocluster cores continues to challenge our understanding of galaxy formation and evolution \citep[see also DRC at $z\,{=}\,4.0$,][]{Oteo2018,Ivison2020}. 
To date, no cosmological simulation or galaxy formation model has successfully predicted their ULIRG overdensity \citep[e.g.,][]{Rennehan2020,Bassini2020,Lim2021,Lim2024,Fukushima2023,Remus2023,Rennehan2024,Tevlin2024}. 
Without relying on additional environmental triggers, the ULIRG duty cycle would need to be significantly extended to account for the coeval ULIRG activity observed in such systems \citep[e.g.,][]{Casey2016}. 
To understand the high SFR density of SPT2349$-$56, it is crucial to investigate cold molecular 
gas ($T_{\rm mol}\,{\lesssim}\,100\rm\,K$), which serves as the fuel of star formation \citep{Kennicutt1998}. 
In particular, large and extended cold gas reservoirs are expected in the circumgalactic and intergalactic media (CGM and IGM) in order to sustain the high levels of star formation in protocluster cores \citep[e.g.,][]{Emonts2016,Emonts2018,Emonts2019,Emonts2023a,Emonts2023b,Ghodsi2024,Chen2024}, which is challenging to capture with high-resolution interferometric observations due to their low surface brightness 
\citep[e.g.,][]{Li2021,Cicone2021,Jones2023}. While ALMA is good at studying the internal properties of individual galaxies, low-resolution data are essential for assessing the global molecular gas reservoir within the system.

In this Letter, we present deep ACA observations of CO(4--3) and 3.2\,mm dust continuum to further explore the integrated molecular gas content of the protocluster system SPT2349$-$56. 
In Section\,\ref{sec:obs} we describe the ALMA/ACA Band-3 and ancillary APEX/FLASH observations and data reduction. 
Section\,\ref{sec:results} presents our estimations of gas mass from three indicators, CO(4--3), 3.2\,mm dust continuum, and {\cii}, as well as the comparison with previous high-resolution ALMA 12-m results. 
A discussion and summary are given in Section\,\ref{sec:discussion}. 
Throughout the paper, we follow the $\Lambda$ cold dark matter cosmology with $H_0\,{=}\,70\,\rm km\,s^{-1}\,Mpc^{-1}$ and $\Omega_m\,{=}\,0.3$, which corresponds to a proper angular scale of 6.88\,kpc/\arcsec\, at $z\,{=}\,4.3$. 

\section{Observations and data reduction}\label{sec:obs} 
\subsection{ALMA observations} \label{subsec:alma} 
Our ACA Band-3 observations of SPT2349$-$56 were obtained as part of the Cycle 10 ALMA program (2023.1.00124.S, PI: S. Chapman). 
We tuned the frequency to place the CO(4--3) line at $z\,{=}\,4.3$ ($\nu_{\rm obs}\approx 86.99\,\rm GHz$) within one of the 1.875\,GHz spectral windows in the lower sideband in the `time division modes' (120 channels with 15.625\,MHz resolution), covering 85.99--87.86\,GHz. 
The program was carried out between 2024 January and 2024 September, with 41 individual execution blocks (EBs) and a total on-source integration time ($t_{\rm int}$) of ${\sim}\,30$ hours. 

We calibrated and reduced the data using the standard calibration script with {\tt CASA} (version\,=\,6.6.1) in the pipeline mode. 
Bad measurement sets with short on-source exposures (${<}\,10$\,min) or high phase errors (${>}\,20^\circ$) were manually discarded after a careful inspection of each QA0 report, remaining 31 EBs with $t_{\rm int}\,{\sim}\,27$ hours. 
Due to the coarse resolution, we performed {\tt CLEAN}ing on continuum subtracted measurement sets using the {\tt CASA} (version\,=\,6.6.5) task {\tt tclean} with `briggs' ({\tt robust\,=\,0.5}) weighting in the `cube' mode with a `hogbom' deconcolver to produce the data cubes for CO(4--3). 
The {\tt CLEAN}ing process contains two steps. 
We first cleaned the data to 4$\sigma$ without a {\tt CLEAN} mask. 
We then masked the high-significance $4\sigma$ channels with adjacent low-significance $2\sigma$ channels \citep[CPROPS mask,][]{Rosolowsky2006} from the 4-$\sigma$ cleaned data cube and cleaned down to 1$\sigma$ following the method described in \citet{Leroy2021}. 
The final product has a synthesized beam of $21.0''\times12.2''$ and reaches a line sensitivity of 0.25\,mJy/beam per channel with a spectral resolution of 54\,km/s. 

High-resolution 12-m observations of CO(4--3) with two pointings were observed in C4 configuration for 2 EBs and in C5 configuration for 5 EBs, which have $t_{\rm int}\,{\sim}\,2.7$ hours for each pointing and reach a maximum recoverable scale (MRS) of 15.1\arcsec. {\cii} observations were conducted in C3 configuration in 3 pointings for 4 EBs ($t_{\rm int}\,{\sim}\,0.97$ hours, MRS\,$\sim$\,6.0\arcsec) and in 6 pointings in C5 configuration for 6 EBs ($t_{\rm int}\,{\sim}\,0.83$ hours, MRS\,$\sim$\,3.0\arcsec) \citep[2017.1.00273.S, PI: S.\,Chapman; 2018.1.00058.S, PI: S.\,Chapman,][]{Hill2020}. The corresponding data cubes were obtained in an identical manner as reported in \citet{Hill2020}, resulting in line sensitivities of 0.07\,mJy/beam for CO(4--3) and 0.22\,mJy/beam for {\cii}, and channel widths of 54\,km/s and 13\,km/s, respectively. The synthesized beams are $1.01''\times0.84''$ and $0.35''\times0.29''$, respectively. 

Continuum images at 3.2\,mm were obtained from line-free channels in `briggs' weighting ({\tt robust\,=\,0.5}) in {\tt mfs} mode with the `hogbom' deconvolver. 
We used a similar strategy to create the {\tt CLEAN}ing mask and cleaned the continuum image down to $1\sigma$. 
The resulting rms level of the whole image is $22.0\,\rm \mu Jy/beam$ with beam size of $15.7''\times 10.6''$. 
We produced the 3.2\,mm high-resolution Cycle 5 Band-3 continuum map in a similar way, using pixels of 0.15\arcsec\, \citep[2017.1.00273, PI: S. Chapman,][]{Hill2020}. 
The synthesized beam size is $0.94''\times0.78''$ and the rms level is $6.0\,\rm \mu Jy/beam$. 

To investigate whether the new ACA results agree with ALMA observations when the resolutions are similar, we produced an ALMA continuum map and CO(4--3) data cube in low-resolutions by applying a $uv$-taper of 10\arcsec, resulting in synthesized beam sizes of $11.0''\times 9.6''$, $9.9''\times 9.4''$ and sensitivity of 0.29\,mJy/beam per channel, 30.2\,$\mu$Jy/beam for CO(4--3) and 3.2\,mm dust continuum, respectively. 
\begin{figure*}[tp!]
    \centering
    \includegraphics[width=0.90\linewidth]{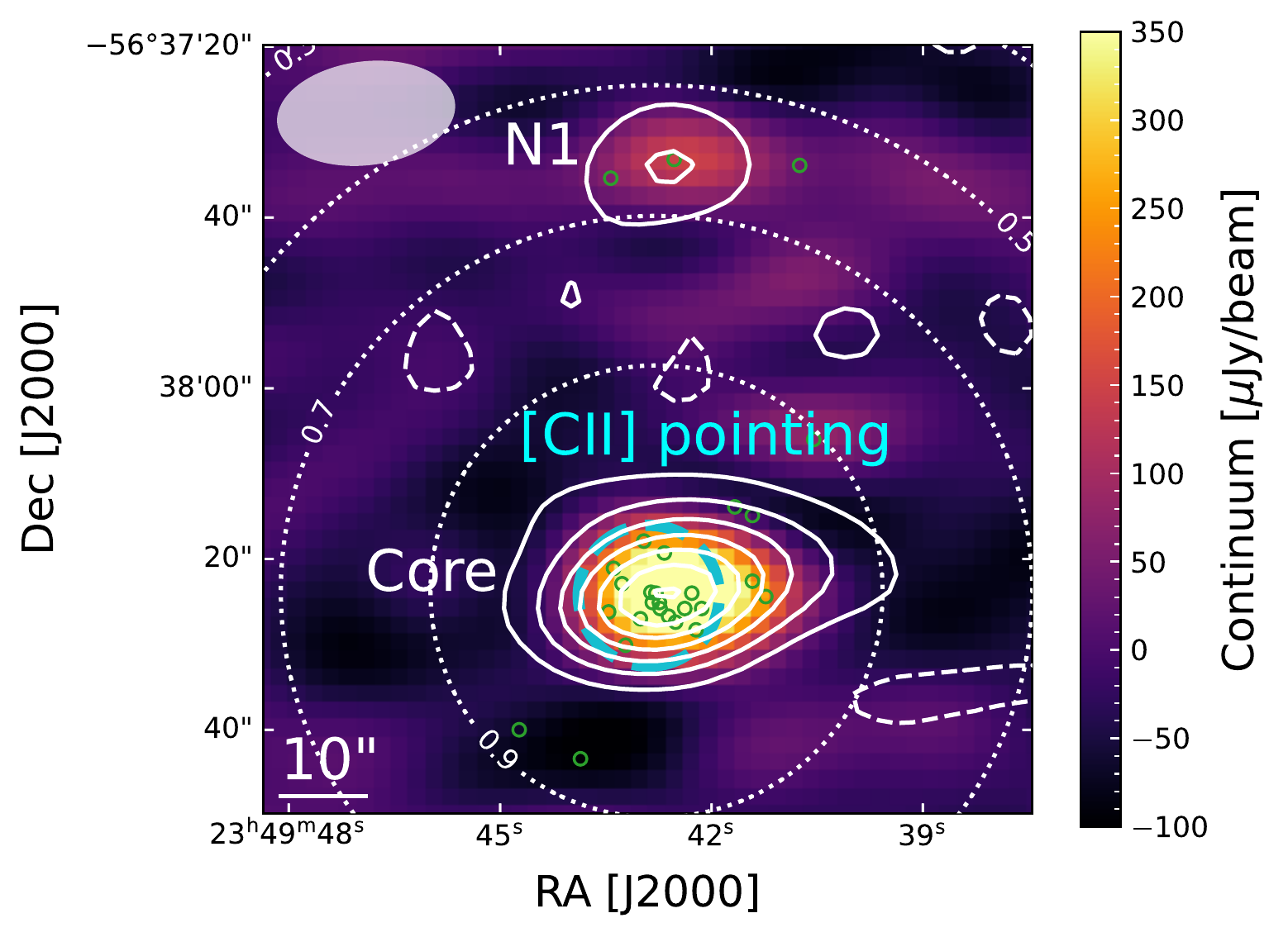}
    \caption{ACA continuum map of the protocluster SPT2349$-$56 with moment-0 contours of CO(4--3) in the velocity window [$-1373\rm\,km/s$,\,$1429\rm\,km/s$] before primary-beam correction. The white solid contours are [3,\,6,\,9,\,12,\,15,\,18,\,21]$\times$\,the rms level of the CO emission from ACA observations. The dashed contours are $-2\times$ the rms level. The white dotted contours indicate [0.3,\,0.5,\,0.7,\,0.9]$\times$ the primary beam response from the continuum map. The cyan dashed circle represents the main beam of the APEX {\cii} observations. Green circles are the locations of all cataloged galaxies from \citet{Hill2020}. We note that the two sources in the vicinity of `N1' are `NL2' and `N3', which are not associated with the structure at $z\,{=}\,4.3$ \citep{Hill2022}.}
    \label{fig:co_map}
\end{figure*}

\subsection{APEX observations}
We also used archival single-dish {\cii} line observations from the APEX FLASH receiver as a part of Max Planck Society observing time \citep{weiss2013,Gullberg2015,Strandet2016,Reuter2020}. 
The observations were conducted under good weather conditions with an average precipitable water vapor of ${\sim}\,0.3\,\rm mm$ and a system temperature of ${\sim}\,300$\,K. 
The total on-source exposure time is about $2.2\,\rm hours$. 
The data were reduced using {\tt \gildas/CLASS}. 
We employed an APEX/FLASH antenna gain\footnote{\url{https://www.apex-telescope.org/telescope/efficiency/}} of $S_\nu/T_{A}^{\ast}=40$~Jy/K for the observed frequency $\nu_\mathrm{obs}=358.4$~GHz \citep{Gullberg2015}. 
We subtracted the baseline with a velocity window of 1600\,km/s and the first order polynomial fitting, leading to a rms level of $27\,\rm mJy$ (0.67\,mK) for a 100\,km/s channel. 
\section{Results} \label{sec:results}
\subsection{ACA observations of CO(4--3)} \label{subsec: aca}
The southern core is only marginally resolved in the ACA observations with the large synthesized beam, allowing a comparison between the integrated CO(4--3) flux and the emission from member galaxies in the protocluster core detected by ALMA (Figure \ref{fig:co_map}). 
The `N1' source in the northern region is relatively isolated, and so a direct comparison to the high-resolution ALMA result is possible. 

First, we used sufficiently large apertures ($30''\,{\times}\,17''$) to obtain the overall ACA spectrum.
For simplicity, we fit the line with a single Gaussian to determine its width and then integrated over ${\pm}\,3\sigma$ from the mean value ([$-1373\,\rm km/s$,\,$1429\,\rm km/s$]) to produce an accurate moment-0 map. 
We varied the size of the elliptical aperture (proportional to the beam shape) and used a curve-of-growth analysis to determine an aperture size which maximizes the total line strength in the central core. 
Similarly, we produced a moment-0 map of `N1' with a velocity range of [$-349\rm\,km/s$,\,$1429\rm\,km/s$]. 
Since `N1' is unresolved in the ACA map, we used the peak-pixel value as the total CO(4--3) line strength. 

To calculate the total line strength in the core region recovered by ALMA 12-m observations, we adopted the same source apertures obtained from the curve-of-growth analysis reported in \citet{Hill2020} for the spectral extraction of the 22 line emitters in the core region. 
We then summed these extracted spectra to obtain an overall CO(4--3) spectrum and integrated the flux density over the same velocity range applied to the ACA spectrum.

\begin{figure}[tp!]
    \centering
    \includegraphics[width=0.93\linewidth]{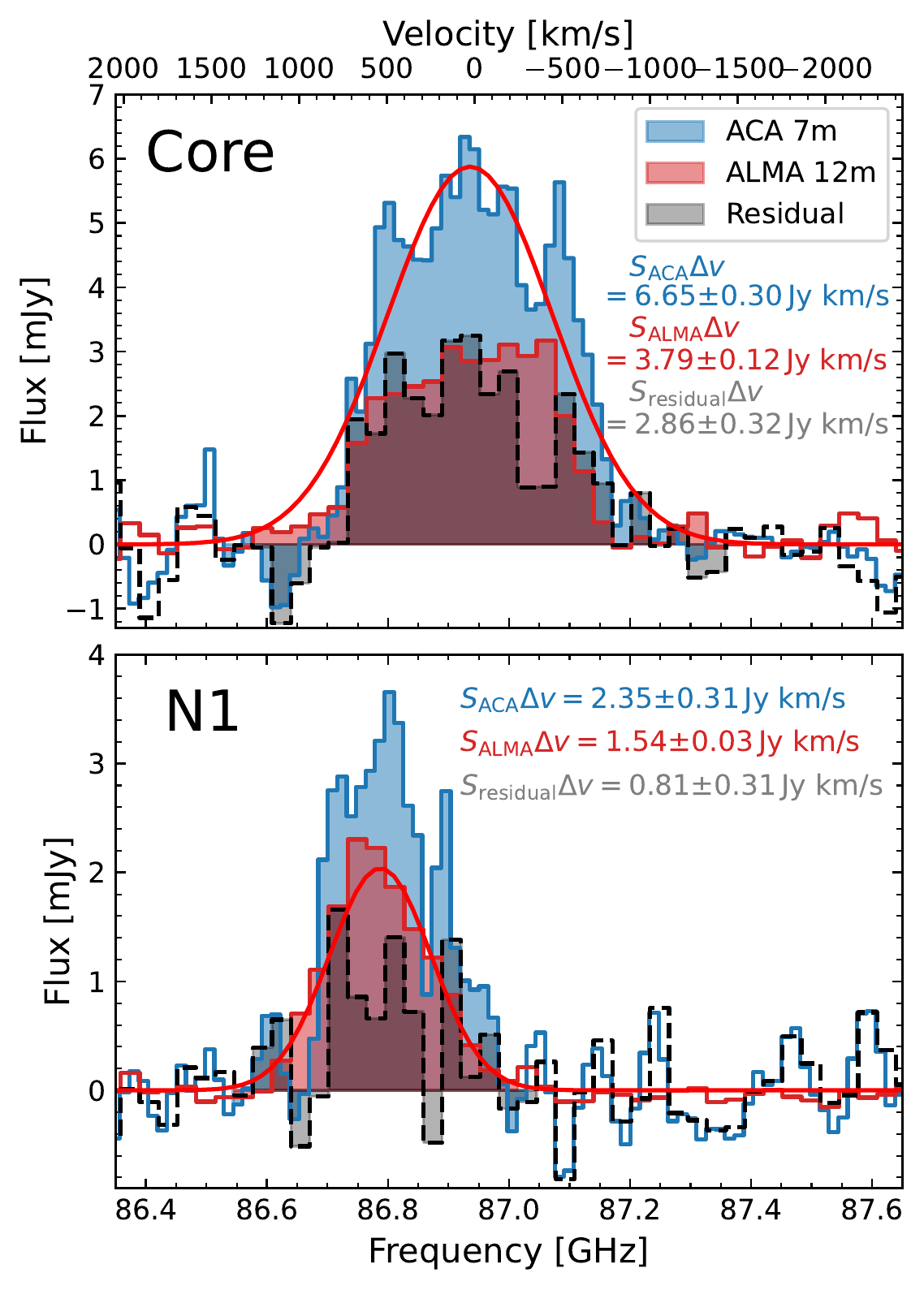}
    \caption{ACA 7-m spectra of CO(4--3) of the southern core and `N1' (blue). The red curves are the best-fit Gaussian profiles for determining the velocity ranges of the moment-0 maps, which are also indicated as the shaded regions. The ALMA 12-m results (obtained by co-adding all individually-detected galaxies) are shown in red, and the differences between the two measurements are shown in black for comparison. Compared to the ALMA 12-m results, the CO(4--3) line strengths are consistently higher in ACA 7-m observations.}
    \label{fig:co_spec}
\end{figure}

The spectra of the core region and `N1' from the ACA 7-m and ALMA 12-m observations are shown in Fig.~\ref{fig:co_spec}, revealing an excess line strength from ACA in both cases (Table~\ref{tab:flux}). 
To ensure the discrepancy is not caused by the potential problem from the absolute flux calibration, we checked their visibilities and corresponding spectra from the tapered ALMA data cube, which also show consistent values (Appendix~\ref{vis}). 
We note that ACA 7-m observations have been reported to systematically underestimate line strengths, which on average causes a 27\% loss in simulations \citep[e.g.,][Appendix C]{Leroy2021}. 
The total CO (4--3) fluxes estimated by ACA observations can be only treated as lower limits. 

We then evaluated the CO line luminosity and molecular gas mass (listed in Table~\ref{tab:flux}) with the following formulas \citep{Solomon2005}:
\begin{equation} \label{solo1}
L_{\mathrm{CO(4-3)}}^{\prime}\,{=}\,3.25{\times}10^7 S_{\mathrm{CO(4-3)}} \Delta v\, \nu_{\text {obs}}^{-2} D_{\mathrm{L}}^2(1{+}z)^{-3} ,
\end{equation}
\begin{equation} \label{solo2}
M_{\rm mol} \,{=}\, \frac{1}{r_{41}} \alpha_{\rm CO} L_{\mathrm{CO(4-3)}}^{\prime},
\end{equation}
where $L_{\mathrm{CO(4-3)}}^{\prime}$ is measured in $\rm K\,km\,s^{-1}\,pc^2$, the velocity-integrated line strength $S_{\mathrm{CO(4-3)}} \Delta v$ is in $\rm Jy\,km\,s^{-1}$, the observed frequency $\nu_{\text {obs}}$ is in GHz, the luminosity distance $D_L$ is in Mpc, and the molecular gas mass $M_{\rm mol}$ is in $M_\odot$. 
We adopted a CO(4--3)-to-CO(1--0) line ratio of $r_{41}\,{=}\,0.60\,{\pm}\,0.05$
and a $L_{\mathrm{CO}}$-to-$M_{\rm mol}$ conversion factor of $\alpha_{\rm CO}\,{=}\,3.8{\pm}0.1\rm\,M_\odot/(K\,km\,s^{-1}\,pc^2)$, which are typical values for high-z SMGs \citep{Spilker2014,Dunne2022}. 
However, the integrated gas masses should only be treated as lower limits, as the extended emission could be emitted from potentially more pristine gas and less extreme galaxies, which would result in a substantially higher $\alpha_{\rm CO}$ and lower $r_{41}$ \citep[e.g.,][]{Tacconi2013,Bolatto2013}.

\begin{figure}[tp!]
    \centering
    \includegraphics[width=0.93\linewidth]{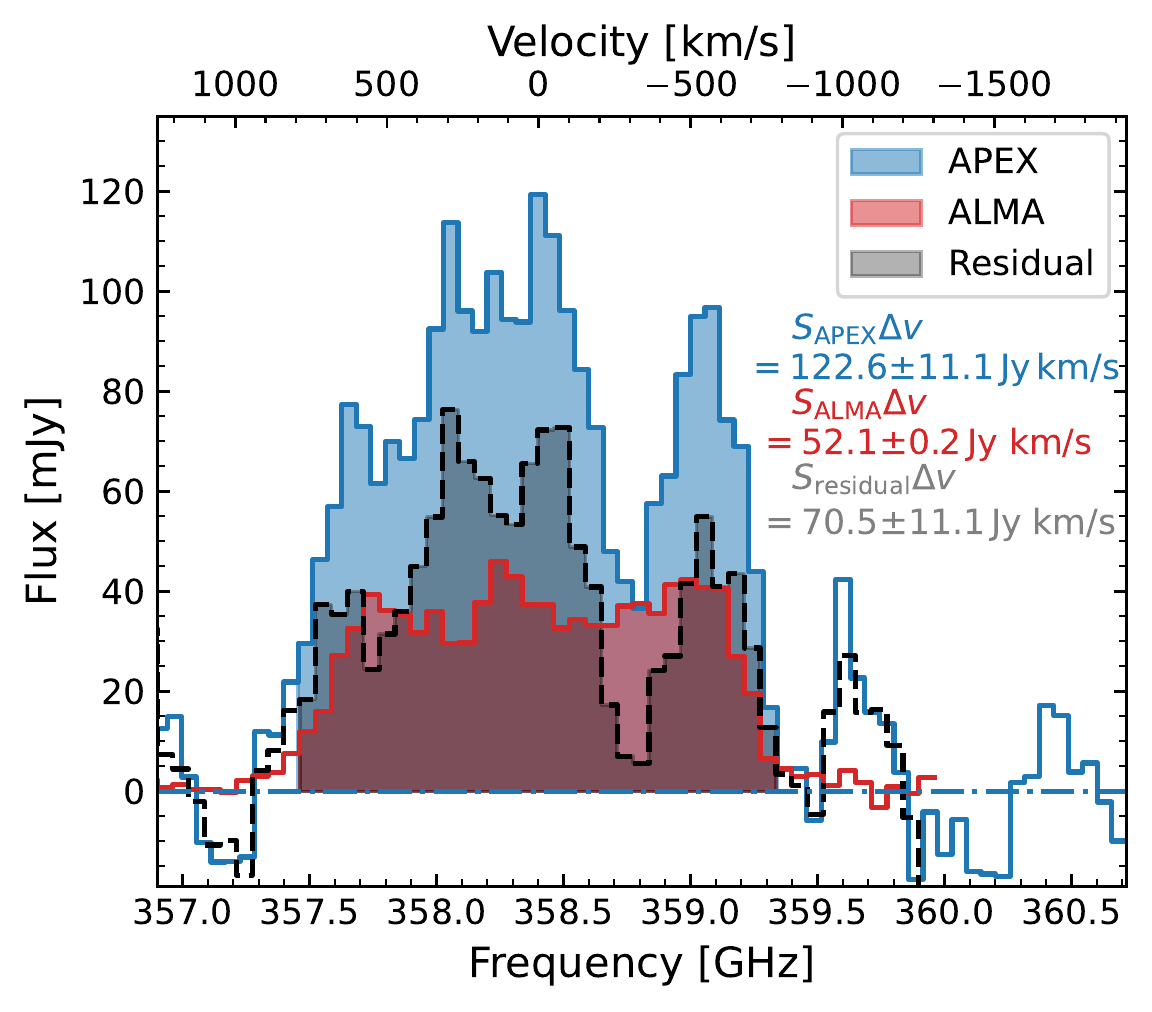}
    \caption{{\cii} spectrum from APEX FLASH observations (blue) compared to the scaled and co-added ALMA spectra of all individually-detected protocluster members within the APEX beam (red). The difference is shown in black. For the purpose of this plot, we slightly smoothed the APEX spectrum with a low-pass Savitzky-Golay filter with a polynomial order of one to suppress the high-frequency noise. Since we use ${\pm}\,800\rm\,km/s$ for the baseline subtraction, only the excluded channels (shaded) were used for calculating the integrated line strength. The excess in {\cii} line strength from the APEX observations supports our ACA result. }
    \label{fig:apex}
\end{figure}
\begin{deluxetable*}{ccccccccc}
\tabletypesize{\small}
\tablewidth{0pt} 
\tablecaption{Comparison of gas mass estimates from ACA/APEX observations to ALMA 12-m observations in the three independent tracers CO, dust continuum, and {\cii}. \label{tab:flux}}
\tablehead{
\colhead{Region} & \colhead{Telescope}& \colhead{$S_{\rm CO(4-3)}\Delta v$} & \colhead{$L'_{\rm CO(4-3)}$} & \colhead{$M_{\rm mol}^{\rm CO}$} & \colhead{$S_{\rm 3.2mm}$} & \colhead{$M_{\rm mol}^{\rm dust}$} & \colhead{$S_{\rm \cii}\Delta v$} & \colhead{$M_{\rm mol}^{\rm\cii}$}\\
& & \colhead{[Jy\,km/s]} & \colhead{[10$^{10}$\,$\rm K\,km\,s^{-1}pc^2$]} & \colhead{[$10^{10}\,\rm M_\odot$]} & \colhead{[$\mu$Jy]} & \colhead{[$10^{10}\,\rm M_\odot$]} & \colhead{[Jy\,km/s]} & \colhead{[$10^{10}\,\rm M_\odot$]}
} 
\colnumbers
\startdata 
    \multirow{3}{*}{Core}& ACA/APEX & 6.65$\pm$0.30 & 29.3$\pm$1.3 & 185.5$\pm$18.2 & 654$\pm$29 & 84.4$\pm$3.7 & 122.6$\pm$11.1 & 209.4$\pm$18.9 \\
    & ALMA & 3.79$\pm$0.12 & 16.7$\pm$0.5 & 105.7.0$\pm$9.8 & 754$\pm$30 & 97.2$\pm$3.9 & 52.1$\pm$0.2 & 89.0$\pm$0.4 \\
    & Excess [\%] & \multicolumn{3}{c}{75.5$\pm$9.6 (CO(4--3))} & \multicolumn{2}{c}{$-13.2{\pm}5.1$ (dust)} & \multicolumn{2}{c}{135.3$\pm$21.3 ({\cii})}\\
    \hline
    \multirow{3}{*}{N1} & ACA & 2.35$\pm$0.31 & 10.4$\pm$1.3 & 65.6$\pm$10.3 & 310$\pm$38 & 39.9$\pm$4.9 & ... & ...\\
    & ALMA & 1.54$\pm$0.03 & 6.77$\pm$0.12 & 42.9$\pm$3.8 & 201$\pm$11 & 25.9$\pm$1.4 & ... & ...\\
    & Excess [\%] & \multicolumn{3}{c}{52.8$\pm$20.0 (CO(4--3))} & \multicolumn{2}{c}{54.2$\pm$20.7 (dust)} & \multicolumn{2}{c}{...}\\
\enddata
\tablecomments{(1): Regions included in this study. (2): The telescope used for the measurement. (3): Integrated CO(4--3) line strength within ${\pm}\,3\sigma$ of the central velocity. The uncertainty is the primary beam-corrected rms value of the moment-0 map for the ACA results, and the overall uncertainty through the error propagation from each channel rms for the APEX/ALMA values. (4) CO(4--3) line luminosity in unit of brightness temperature derived from Eq.\,\ref{solo1}. (5) Gas mass estimated from Eq.\,\ref{solo2}. The error includes uncertainties from the measurements and from the chosen $\alpha_{\rm CO}$ and $r_{41}$ values. (6) Total 3.2\,mm continuum flux density from Table~\ref{tab:3mm_cont}. (7) Gas mass estimated from Eq.\,\ref{sco1}. (8) Integrated {\cii} emission line strength within $\pm$\,800\,km/s of the central velocity, which is defined by the velocity range for the baseline subtraction. The quoted ALMA values have been corrected for the APEX primary beam response with a FWHM of 16.5\arcsec. We used the ${<}\,{-}800\rm\,km/s$ and ${>}\,800\rm\,km/s$ channels to fit the baseline with a first order polynomial for the APEX result. Both errors were obtained from the error propagation. (9) Gas estimates from {\cii} line emission with Eq.\,\ref{zan1}. }
\end{deluxetable*}

\subsection{ACA observations of dust continuum} \label{sec:cont}
To investigate whether the low-resolution observations result in systematically higher values across different tracers, we further compared the long-wavelength dust continuum \citep[often used as a tracer of gas mass, e.g.,][]{Scoville2016,Scoville2017,Scoville2023,Dunne2022} in the core region and in `N1' using the same ACA and 12-m data. 
As the 3.2\,mm continua haven't been significantly detected in most protocluster members, we summed the flux densities of sources with a $4\sigma$ detection and scaled the undetected sources from their 850\,$\mu$m flux densities to avoid the large uncertainty.
We tested three separate methods; we recorded the continua measured from the peak pixel values, we integrated the surface brightness within Kron apertures ({\tt k\,=\,2.5, Rmin\,=\,3.5}) using {\tt SourceExtractor} \citep[version 2.28.0,][]{Bertin1996}, and we modeled the image using a single-component 2D Gaussian using the {\tt CASA} task {\tt imfit} (Table~\ref{tab:3mm_cont}). 
Considering the complex extended structure of the integrated core galaxies versus the compact continuum source in `N1', we adopted the kron aperture flux and the peak value for the 3.2\,mm continuum in the core region and `N1', respectively. 

We ran {\tt SourceExtractor} again in a similar manner to estimate the 3\,mm continuum recovered by the ALMA 12-m data. 
We recorded the three independent measurements for sources with significant detections (${>}\,4\sigma$). 
We carefully inspected the kron apertures and image cuts of each source and determined their flux densities based on their compactness and morphologies. 
We note that only seven sources (corresponding to 10 of the cataloged galaxies due to blending) that lie within the ACA 2$\sigma$ continuum contour of the core region have been detected by the 12-m data, leaving 12 confirmed protocluster members remaining undetected due to the faintness of their long-wavelength continua. 
To account for the continuum fluxes from these undetected sources, we estimated their expected 3\,mm continuum levels based on their $850\rm\,\mu m$ flux densities with the scaling factor $S_{850\rm\mu m}/S_{\rm 3.2mm}$ obtained from the detected sources. 

As shown in Table~\ref{tab:3mm_cont}, the `N1' source has a somewhat ($\approx$\,50\%/2.6$\sigma$) higher 3.2\,mm continuum with ACA than we measured with the 12-m data, which could be caused by the contribution from faint surrounding galaxies (e.g., interlopers `NL2' and `N3') or the large uncertainty of `N1', which is away from the phase center of the ACA pointing. 
This is supported by the fact that `N1' only shows a mild increase in continuum flux when the ALMA continuum map is tapered (Appendix\,\ref{vis}), which suggests that its dust continuum could be overestimated from the ACA continuum map. 
In contrast, compared to the sum of the 12-m source measurements, the core region shows a comparable or even weaker 3.2\,mm continuum (${\approx}\,-$10\%/$-2.6\sigma$) with ACA. 
However, the differences from ACA to 12-m in both the core and `N1' are still within $3\sigma$ uncertainties. 

We note that the lower continuum level in the core region measured with the ACA may be a result of a smaller contribution from noise boosting for the unresolved core, or contamination from the SZ effect in the low-resolution data \citep[][in prep]{Sunyaev1980,Zhou2024}. 
However, the contrast with the large excess in CO(4--3) found by ACA suggests the difference between CO(4--3) and continuum is most likely astrophysical in origin, which will be discussed in Section\,\ref{sec:discussion}. 

We used the following formula to estimate the gas mass from the 3.2\,mm dust continuum in units of $10^{10}\,M_{\odot}$\citep{Scoville2016,Scoville2017,Scoville2023}:
\begin{equation} \label{sco1}
\begin{aligned}
M_{\mathrm{mol}}=\, & 1.78\, S_{\text {obs }} (1+\mathrm{z})^{-(\beta+3)}\left(\frac{\nu_{850}}{\nu_{\mathrm{obs}}}\right)^{\beta+2}D_L^2,
\end{aligned}
\end{equation}
where $S_{\text {obs}}$ is the observed continuum flux density in the unit of mJy, $\beta$ is the dust emissivity index, ${\nu_{850}}$ and $\nu_{\mathrm{obs}}$ are the frequencies at $850\rm\,\mu m$ and at the observed wavelength, and the luminosity distance $D_L$ is in Gpc. 
We used the typical emissivity index of $\beta\,{=}\,2$ for the calculation. 
The estimated continuum-derived gas masses are also shown in Table~\ref{tab:flux} for comparison.  

\begin{deluxetable*}{ccccccccccc}
\tabletypesize{\small}
\tablewidth{0pt} 
\tablecaption{3.2\,mm source catalog from the ACA and ALMA observations. See Section~\ref{sec:cont} for details.\label{tab:3mm_cont}}
\tablehead{
    \colhead{ID} & \colhead{RA} & \colhead{Dec} & \colhead{Distance} & \colhead{S/N} & \colhead{PB} & \colhead{S$_{\rm peak}$} & \colhead{S$_{\rm kron}$} & \colhead{S$_{\rm fit}$} & \colhead{S$_{\rm best}$} & \colhead{Counterpart} \\  
    & \colhead{[deg]} & \colhead{[deg]} & \colhead{[arcsec]} & & & \colhead{[$\mu$Jy]} & \colhead{[$\mu$Jy]} & \colhead{[$\mu$Jy]} & \colhead{[$\mu$Jy]} & }
\colnumbers
\startdata 
1$^*$ & 357.4278 & $-$56.6387 & 4.6 & 18.3 & 1.00 & 110$\pm$6 & 149$\pm$12 &142$\pm$12&149$\pm$12& C1(A)\\
2 & 357.4225 & $-$56.6396 & 9.5 & 12.9 & 0.92 & 85$\pm$7 & 95$\pm$13 &98$\pm$8&98$\pm$8& C4(D)\\
3$^*$ & 357.4284 & $-$56.6403 & 2.6 & 13.9 & 0.99 & 84$\pm$6 & 106$\pm$10 &119$\pm$15&106$\pm$10 & C6,C13,C16(CG)\\
4$^*$ & 357.4282 & $-$56.6400 & 1.9 & 19.6 & 0.99 & 119$\pm$6 & 137$\pm$12 &155$\pm$17& 137$\pm$12 & C3,C12(B)\\
5$^*$ & 357.4218 & $-$56.6402 & 10.8 & 6.7 & 0.89 & 45$\pm$7 & 87$\pm$16 &70$\pm$16&87$\pm$16& C8(E)\\
6 & 357.4256 & $-$56.6405 & 3.8 & 9.5 & 0.97 & 59$\pm$6 & 65$\pm$14 & 61$\pm$8 &59$\pm$6& C5(F)\\
7 & 357.4259 & $-$56.6411 & 4.9 & 4.4 & 0.95 & 28$\pm$6 & 31$\pm$15 &28$\pm$7&28$\pm$6& C9(I)\\
rest$^{**}$ &\multicolumn{4}{c}{$S_{\rm 850,total}\,{=}\,4.67\,{\pm}\,0.32\,\rm mJy$} & \multicolumn{4}{c}{$f_{\rm cor}\,{=}\,52.0{\pm}4.5$} & 90$\pm$10 &...\\
\hline
Core (12-m) &...&...&...&...&...&...&...&...& 754$\pm$30 &...\\
Core$^*$ (ACA) &357.4272 &$-$56.6399&...&22.6 &1.00 & 499$\pm$22 & 654$\pm$29 & 712$\pm$28 &  654$\pm$29&...\\
\hline
N1$^*$ (12-m) & 357.4272 & $-$56.6259 & 2.0&24.1&1.00&169$\pm$7&201$\pm$11&180$\pm$10 & 201$\pm$11&...\\
N1 (ACA) & 357.4262 & $-$56.6261 & ... & 8.1 & 0.57 & 310$\pm$38 &223$\pm$35 &238$\pm$52 & 310$\pm$38&...\\
\enddata
        \tablecomments{(1): Source ID. (2)-(3): Source coordinate of the peak in the continuum map. (4): The distance from the peak position of the ALMA sources to the peak position of the ACA source. (5): The signal-to-noise ratio (S/N) of the peak flux density obtained in the continuum map before primary-beam correction. (6): The primary beam response of the ACA/ALMA map. (7): The peak flux density and corresponding rms level after the primary-beam correction. (8): The total flux within a Kron aperture determined by {\tt SourceExtractor} with the primary-beam correction. The uncertainty is obtained from the standard deviation of 10,000 random kron apertures placed on the continuum map before primary beam correction and then corrected by the beam response. (9): The primary beam-corrected flux density estimated with a single-component 2D Gaussian fit using {\tt imfit} in CASA. (10): Our best estimate of the continuum flux density after visual inspections. We adopted the peak value for the unresolved sources, the model flux density for the resolved sources that can be well-fit by a 2D Gaussian profile, and the aperture flux density for the resolved sources with a complex morphology. (11): The corresponding counterpart identified from \citet{Hill2020} and \citet{Miller2018}.\\
        $^*$: The source is resolved and displays a complex morphology. We chose the aperture flux as the best result. \\
        $^{**}$: Because a substantial number of sources remain undetected, we calculated the average $S_{\rm 850\mu m}/S_{\rm 3.2mm}$ ratio $f_{\rm cor}$ from the seven detected sources and estimated the overall scaled 3.2\,mm flux from the undetected sources.}
\end{deluxetable*}

\subsection{APEX observations of [CII]} \label{subsec:apex}
Considering the discrepancies seen in ACA versus 12-m through two gas tracers, CO(4--3) and long-wavelength dust continuum, we also utilized archival APEX observations to see if an excess exists in {\cii}, also an indicator of gas mass \citep[e.g.,][]{Carilli2013,Zanella2018}. 
Fig.~\ref{fig:apex} shows the {\cii} spectrum of the protocluster core, where 
the pointing center and beam are indicated on Fig.~\ref{fig:co_map}. 
To compare the APEX spectrum to the integrated {\cii} emission from individual sources detected with ALMA, the APEX beam response must be taken into account. 
We first obtained the spectra of the individual ALMA {\cii} sources in a manner identical to that described in Section~\ref{subsec: aca}. 
We then calculated their distance to the APEX beam center and corrected the corresponding 12-m {\cii} spectra for the APEX beam response. 
For simplicity, we used a Gaussian function to approximate the APEX beam for each ALMA source attenuation and treated the galaxies as point sources. 
The integrated {\cii} line strength was obtained within a range of $\pm$800\,km/s from the line center, which is defined by the boundary of the baseline subtraction of the APEX spectrum. 

We note that the pointing on APEX introduces some uncertainties into the beam center. 
Considering ${\approx}\,200$ dumps are included in $\sim$\,20 hours of observations (including calibrations), some effective beam size uncertainty may be introduced. 
We thus considered a range of beam sizes from 16.5$''$ to 20$''$ for the correction. 
Compared to the standard FWHM of 16.5\arcsec, the largest beam size (20\arcsec) results in a 9\% difference of the integrated ALMA {\cii} flux, which is small compared to the calibration uncertainty. 
Therefore, we adopted the observatory-suggested beam size of 16.5\arcsec\, for our comparison (Table~\ref{tab:flux}). 

We then calculated the {\cii} luminosity and the gas mass from the following relations \citep{Solomon2005,Zanella2018}:
\begin{equation} \label{zan1}
L_{\mathrm{\cii}}\,{=}\,1.04 \times 10^{-3} S_{\mathrm{\cii}} \Delta v \,\nu_{\text {rest}}(1+z)^{-1} D_{\mathrm{L}}^2,
\end{equation}
\begin{equation}
M_{\mathrm{mol}} \,{=}\, \alpha_{\rm \cii}L_{\rm \cii},
\end{equation}
where the {\cii} line luminosity $L_{\mathrm{\cii}}$ is in units of $L_\odot$, the velocity-integrated line strength $S_{\mathrm{\cii}} \Delta v$ is in $\rm Jy\,km\,s^{-1}$, the luminosity distance $D_L$ is in Mpc, and the molecular gas mass $M_{mol}$ is in $M_\odot$. 
For the derived gas mass in Table~\ref{tab:flux} we used the standard {\cii}-to-$M_{\rm mol}$ conversion factor $\alpha_{\rm \cii}\,{=}\,30\,M_\odot/L_\odot$ \citep[e.g.,][]{Zanella2018,Decarli2022}. We note that due to the APEX beam response, the measured {\cii} flux is underestimated, which is only used for comparison purposes. 

\section{Discussion and Summary} \label{sec:discussion}
In Section~\ref{sec:results} we compared the emissions from three different gas tracers, CO(4--3), dust continuum at 3.2\,mm, and {\cii}, measured from high-resolution ALMA 12-m observations and low-resolution ACA 7-m and APEX/FLASH observations. 
We found that compared to the dust continuum, both CO(4--3) and {\cii} show a significant excess in the low-resolution data. 
As the flux loss from the aperture measurement is only up to about $10\%$ \citep{Hill2020}, the finite aperture size is less likely to miss such a large amount of the emission. 
We also carefully investigated CO(4--3) in the $uv$-tapered ALMA data cube and the $uv$-plane and concluded that the discrepancies were not caused by the absolute calibration uncertainties (Appendix~\ref{vis}), suggesting a significant fraction of emission should not arise from the ISM of the confirmed protocluster members. 

It is of interest to understand the possible origins of the additional gas captured by the low-resolution data. 
First, some faint line emitters might be missed due to the limited sensitivity. 
\citet{Hill2020} adopted a strict threshold for {\cii}-emitter selection (S/N\,$\geq$\,6.2) to ensure a high fidelity. 
A missing population of faint emitters might contribute a considerable amount of {\cii} and CO(4--3) emission (Sulzenauer et al.\,in prep). 
However, additional ${>}\,100$ faint line emitters with $F_{\rm CO(4-3)}\,{\gtrsim}\,0.03\rm\,Jy\,km\,s^{-1}$ would be required to account for the missing CO(4--3) emission in the central 100-kpc radius region. 
Such a number of faint emitters cannot be accommodated by the current luminosity function in SPT2349$-$56, where the flat faint end is reasonably well constrained \citep{Hill2020}. 

Alternatively, the discrepancy could be due to the extended CGM around protocluster galaxies \citep[e.g.,][]{Emonts2016,Emonts2023b,Chen2024,Solimano2024}. 
Literature studies of CO(1--0) in high-$z$ galaxies indicate ubiquitous extended CO(1--0) halos, which suggests that a bulk of molecular gas (up to 80\%) arises from clumpy dense clouds outside of star-forming regions \citep[e.g.,][]{ivison2011,Frias2023,Chen2024,Rybak2024}. 
It is possible that the molecular CGM arises from outflows caused by stellar or AGN feedback \citep[e.g.,][]{weiss2001,Walter2002,Lutz2020,Levy2021,Montoya2024}, which was trapped by the deep potential well of SPT2349$-$56 and accelerates the baryon cycle process \citep[e.g.,][]{Ginolfi2020,Faucher2023,Thompson2024}.

The `negative {\tt CLEAN}ing bowls' in high-resolution interferometric data is a common artifact caused by inadequate $uv$-sampling when the extended emission is over-resolved, which appear as negative regions spatially and spectrally around bright emission lines in deconvolved data cubes. 
This feature has been identified from the {\cii} spectra of protocluster members in the ALMA 12-m observations \citep[see][Appendix~A]{Hill2020}, suggesting missing extended components in the current high-resolution data cube \citep[e.g.,][Sulzenauer et al.\,in prep]{Leroy2021,Czekala2021}. 
A significant flux loss is expected due to the inadequate {\tt CLEAN}ing, which is usually pronounced in more diffuse emission and the higher resolution data. 
It naturally explains the discrepancies from the three different gas tracers. 
An accurate {\tt CLEAN} mask with a perfect {\tt CLEAN}ing is needed to recover the flux loss. 

Another possible origin of the missing flux could be the very extended gas reservoir, permeating the protocluster core before virialized heating occurs \citep[i.e., the `proto-intracluster medium' or proto-ICM, e.g.,][]{Dong2023,Heintz2024}, which would have low surface brightness and may reach the MRS ($\sim$\,15\arcsec) of previous ALMA configurations. 
Some pre-processing processes such as ram pressure stripping, tidal interaction, and harassment are expected in such a crowded field, causing a substantial amount of gas to be stripped but remain in the protocluster \citep[e.g.,][]{Alberts2022,Tevlin2024}. 
Interestingly, despite the discrepancy in derived gas mass seen between the different tracers (Table \ref{tab:flux}), the gas masses derived from CO(4--3) and from the dust continuum show a good agreement in the ALMA 12-m data, which indicates that the CO(4--3) and dust continuum emission captured by the high-resolution interferometric observations trace the same gas component. 
The consistent dust continuum but higher CO(4--3) emission in the ACA observations imply a very high gas-to-dust ratio of this extra gas, which is difficult to reach by the ISM but plausible from a pre-heated intracluster medium \citep[e.g.,][]{Giard2008}. 

Indeed, the deeper {\cii} observations do not only discover additional faint {\cii} emitters, but reveal extended `{\cii} streamers', suggesting that a fraction of cold gas is not in the form of the ISM or the CGM \citep[][Sulzenauer et al.\,in prep]{Hill2020}. But the `{\cii} streamers' only contribute to ${\sim}\,5$\% of the total {\cii} emission in current ALMA data, which cannot fully explain the excess flux (Sulzenauer et al.\,in prep). 
In addition, the consistently high {\ci}(2--1)/{\ci}(1--0) ratios of the protocluster members in SPT2349$-$56 also point to a scenario where cold gas might have been stripped and remains in the form of the CGM or proto-ICM \citep{Hughes2024}. 
In this case, the missing flux can be recovered with a heavy $uv$-tapering when its corresponding $uv$-distance is larger than the shortest baseline. 
For emission arises from even a larger scale, the compact configurations of ALMA and the additional observations with the ACA 7-m and even the total-power (TP) array might be needed to add more short spacings to recover the large-scale missing flux. 

The end point of this additional extended gas is unclear. It could be a reservoir replenishing the star formation fuel, or the early stage of the emergent proto-ICM before heating during virialization. 
If the gas can be efficiently accreted onto the protocluster galaxies, it could provide a sustained fuel, counteracting the rapid gas depletion caused by high SFRs. 
By adopting the total SFR estimated by \citet{Hill2020} and our updated gas mass for the core region, we re-evaluated the gas-depletion time scale. 
The overall depletion time becomes ${\gtrsim}$\,400\,Myr, which significantly relaxes the synchronization requirements for ${>}\,10$ coeval ULIRGs in SPT2349$-$56 \citep[][see Fig.\,4]{Casey2016}. 
If there is no further accretion from the cosmic web, the high SFRs can still be maintained until $z\,{\sim}\,3$, when the quenching is expected to take place. 
Beyond its potential role in fueling star formation, this gas could also represent the cold phase of the emergent ICM. 
During this transition phase, the cold gas could begin to accumulate and interact within the protocluster environment, eventually heating itself through gravitational relaxation while evolving into a mature galaxy cluster \citep[e.g.,][]{Remus2023}.

In summary, we have presented deep ACA Band-3 observations of CO(4--3) and 3.2\,mm dust continuum in the protocluster SPT2349$-$56. 
Compared to the high-resolution ALMA 12-m observations, the ACA observations show a $75\%$ excess in CO(4--3), but a smaller difference in the 3.2\,mm continuum flux. 
We also utilized archival APEX observations to show that {\cii} emission is also $135$\% higher in the single-dish data compared to the ALMA 12-m data. 

Our analysis shows that the discrepancies in the flux excess from these gas tracers are not likely to be caused by missing faint galaxies, since the faint end of the luminosity function would need to be far steeper than previously constrained. 
Instead, the origin of the missing flux could arise from the extended emission, either from the large-scale CGM or the more diffused cold `proto-ICM' before virialized heating. 
The observed excess gas suggests the buildup of the nascent ICM, or an extended gas reservoir that can prolong the depletion time and enable the high SFRs simultaneously observed across $>$\,10 ULIRGs in SPT2349$-$56. 
Strategies such as deep {\tt CLEAN}ing, $uv$-tapering, or adding more short spacings from the ALMA compact configurations or the ACA and TP arrays might help recover the missing flux in high-resolution ALMA observations of systems similar to SPT2349$-$56. 

\begin{acknowledgments}
We thank John E. Carlstrom, Kevin Harrington, Daizhong Liu, Padelis
Papadopoulos, and Douglas Rennehan for useful discussions on this work. 
This research used the Canadian Advanced Network For Astronomy Research (CANFAR) operated in partnership by the Canadian Astronomy Data Centre and The Digital Research Alliance of Canada with support from the National Research Council of Canada the Canadian Space Agency, CANARIE and the Canadian Foundation for Innovation. 
This paper makes use of the following ALMA data: ADS/JAO.ALMA\#2017.1.00273.S, ADS/JAO.ALMA\#2018.1.00058.S, and ADS/JAO.ALMA\#2023.1.00124.S. 
ALMA is a partnership of ESO (representing its member states), NSF (USA) and NINS (Japan), together with NRC (Canada), MOST and ASIAA (Taiwan), and KASI (Republic of Korea), in cooperation with the Republic of Chile. 
The Joint ALMA Observatory is operated by ESO, AUI/NRAO and NAOJ. 
The National Radio Astronomy Observatory is a facility of the National Science Foundation operated under cooperative agreement by Associated Universities, Inc. 
This publication is based on data acquired with the Atacama Pathfinder Experiment (APEX). APEX is a collaboration between the Max-Planck-Institut fur Radioastronomie, the European Southern Observatory, and the Onsala Space Observatory. 
The SPT is supported by the National Science Foundation through grant PLR-1248097, with partial support through PHY-1125897, the Kavli Foundation, and the Gordon and Betty Moore Foundation grant GBMF 947. 
DZ, SCC, RH, and GCPW acknowledge support from NSERC-6740. 
MA acknowledges support from ANID Basal Project FB210003 and and ANID MILENIO NCN2024\_112. 
MS was financially supported by Becas-ANID scholarship \#21221511, and also acknowledges support from ANID BASAL project FB210003. 
\end{acknowledgments}
\vspace{5mm}
\facilities{ALMA Band-3 \citep{band3}, APEX/FLASH \citep{flash}}
\software{astropy \citep{astropy2022}, spectral\_cube \citep{spectralcube}, \gildas/CLASS \citep{gildas}, numpy \citep{numpy}, scipy \citep{2020SciPy-NMeth}, matplotlib \citep{matplotlib}, CASA \citep{CASA}, uvplot \citep{uvplot}, {\tt SourceExtractor} \citep{Bertin1996}}

\appendix
\section{Visibility Analysis}\label{vis}
\begin{figure}[htp!]
     \includegraphics[width=0.44\linewidth]{uv_cov.png}
     \includegraphics[width=0.55\linewidth]{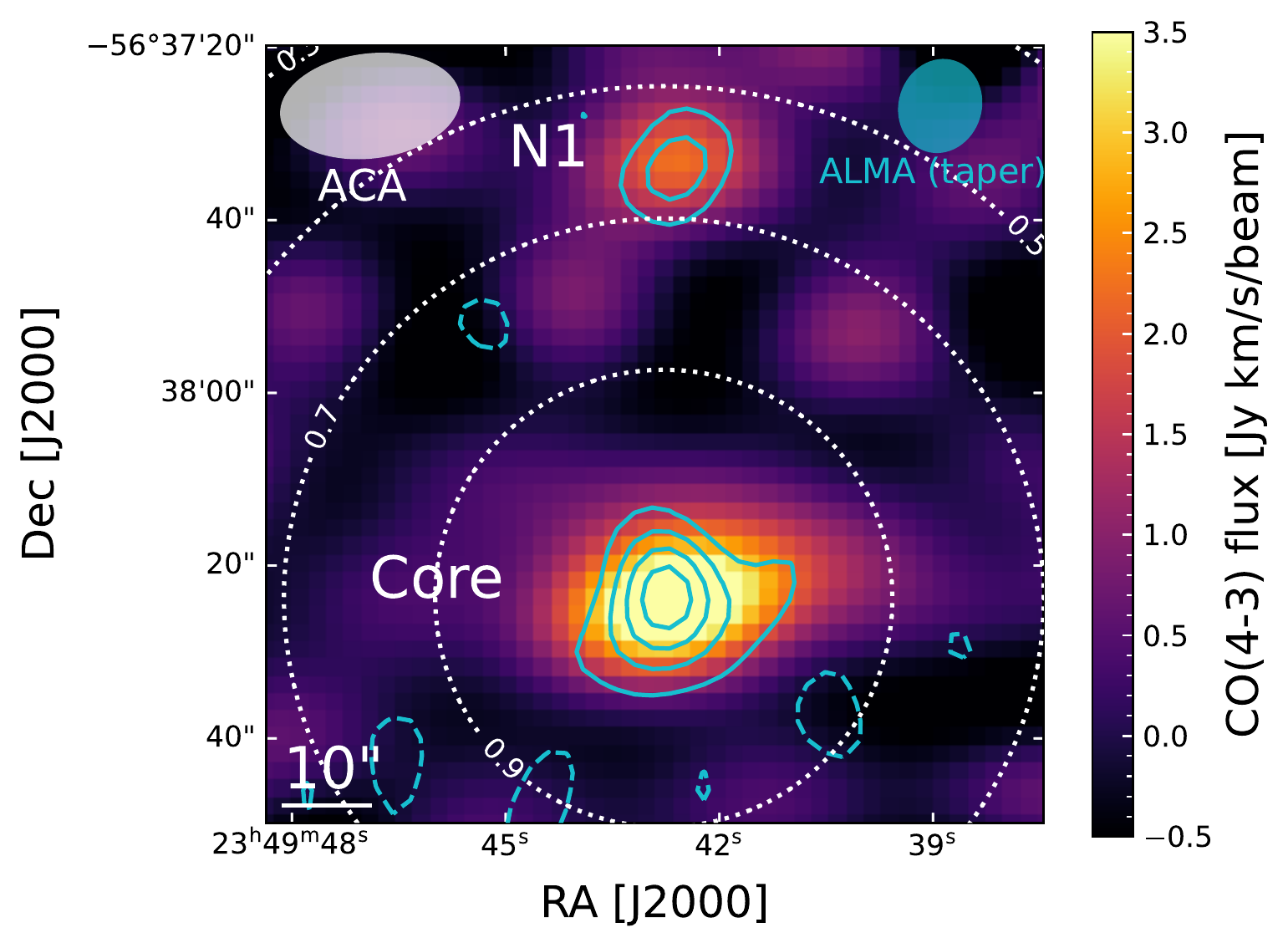}
     \caption{{\it Left:} $uv$-coverage of ACA (blue) and ALMA (red) observations. {\it Right:} ACA CO(4--3) moment-0 map overlaid with tapered ALMA CO(4--3) moment-0 contours. Details and line styles are similar to Fig.~\ref{fig:co_map}. The synthesized beams of ACA (white) and tapered ALMA (blue) are shown on the top left and top right. }
     \label{fig:taper}
\end{figure}

To compare if the difference seen in ACA observations is caused by the absolute calibration uncertainties, we also measured the integrated CO(4--3) flux and 3.2\,mm continuum flux density of the southern core directly in the visibility plane, which avoids potential issues from the over-resolved flux with the {\tt CLEAN}ing. The $uv$-coverage of ACA and ALMA data are shown in Fig.\,\ref{fig:taper} (left panel).

\begin{figure*}[htp!]
     \centering
     \includegraphics[width=0.48\linewidth]{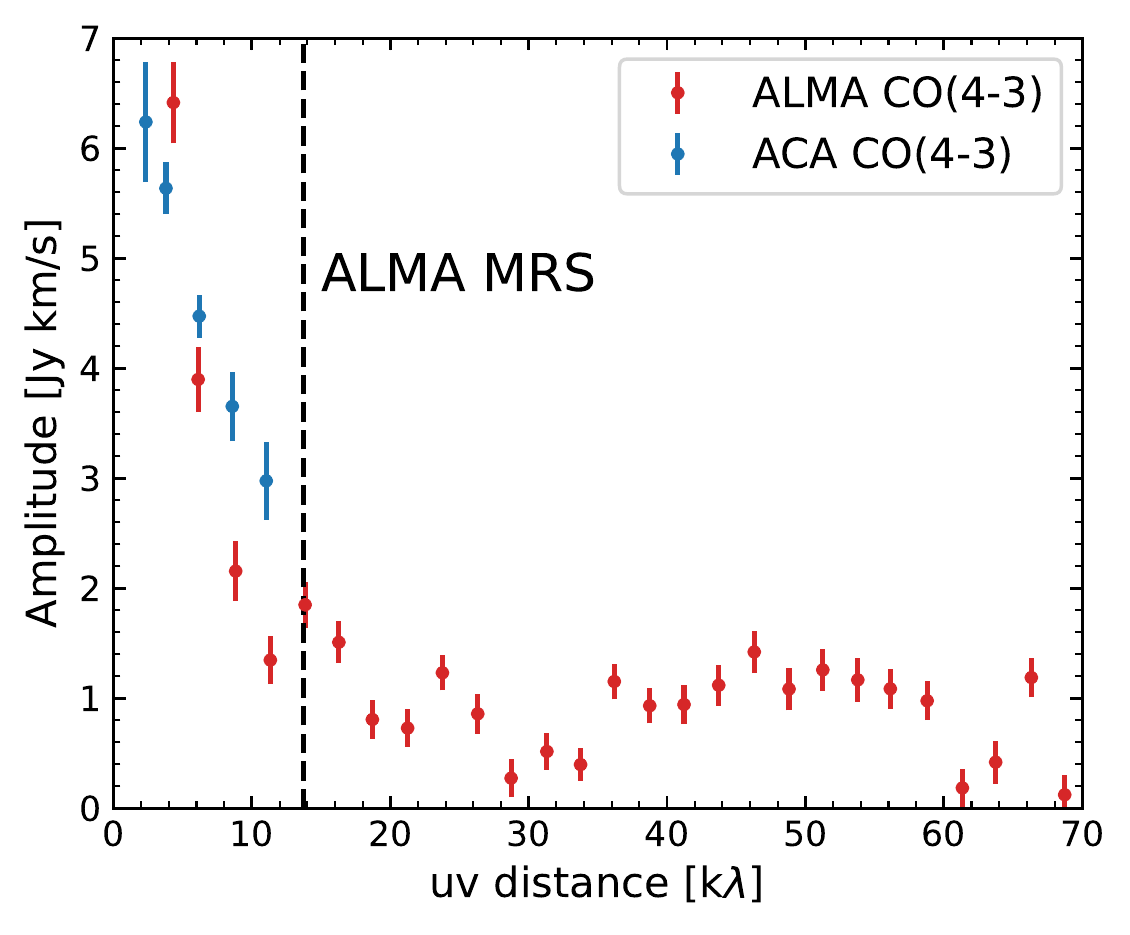}
     \includegraphics[width=0.5\linewidth]{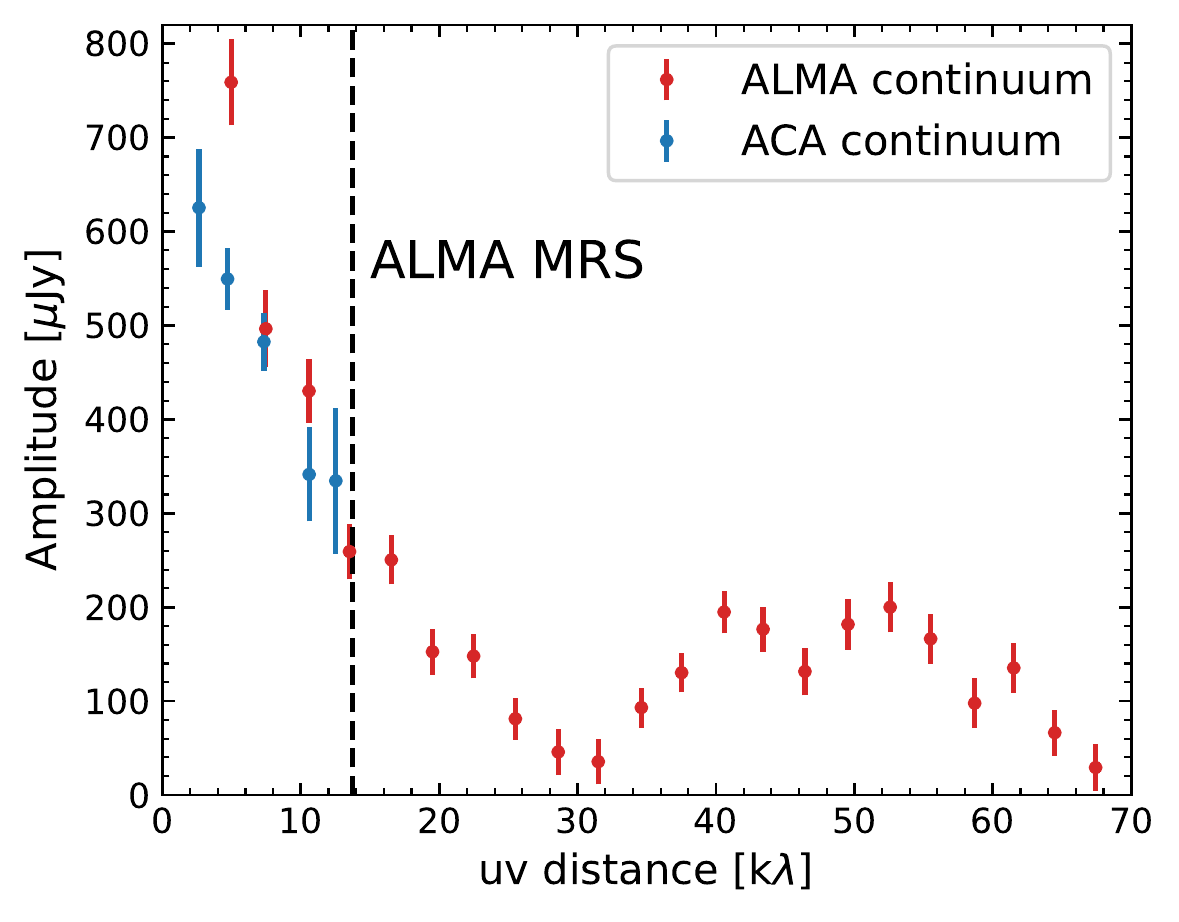}
     \caption{Amplitudes of CO(4--3) and 3.2\,mm dust continuum of the southern core from the ALMA (red) and ACA (blue) observations in the visibility plane. The MRS of the ALMA 12-m observations is shown as black dashed lines. Visibilities show a good agreement at similar $uv$ distance, which suggests that CO(4--3) excess in ACA observations is not caused by the absolute calibration uncertainty.}
     \label{fig:vis}
\end{figure*}

For the CO(4--3) emission, we used the continuum-subtracted measurement sets obtained from CASA task {\tt uvcontsub} and spectrally averaged them over the velocity range within 1.2$\times$FWHM of the line width with CASA task {\tt split} and compensated the flux loss measured in the ACA spectrum to maximize the S/N \citep{Novak2020}.
We then scaled the visibilities by the corresponding velocity range and calculated weight-average visibility in each $uv$-distance bin. 
For the comparison of the continuum flux densities, we flagged all channels associated with CO emission and averaged all channels in each spectral window to obtain visibilities of the dust continuum. 
The $\sim$\,2{\arcsec} offset in the phase centers of two observations was corrected with the {\tt uvplot} function {\tt apply\_phase}.
Because an accurate multi-component fitting is beyond the scope of this work, we only compared the vector-averaged amplitudes of the visibilities of ACA and ALMA observations obtained from {\tt uvplot}. 
As shown in Fig.\,\ref{fig:vis}, the visibility amplitudes of both CO(4--3) and 3.2\,mm continuum are consistent at a similar $uv$-distance, which suggests that the discrepancy should not be caused by the flux scale from different amplitude calibrations.

\begin{figure*}[htp!]
     \centering
     \includegraphics[width=0.95\linewidth]{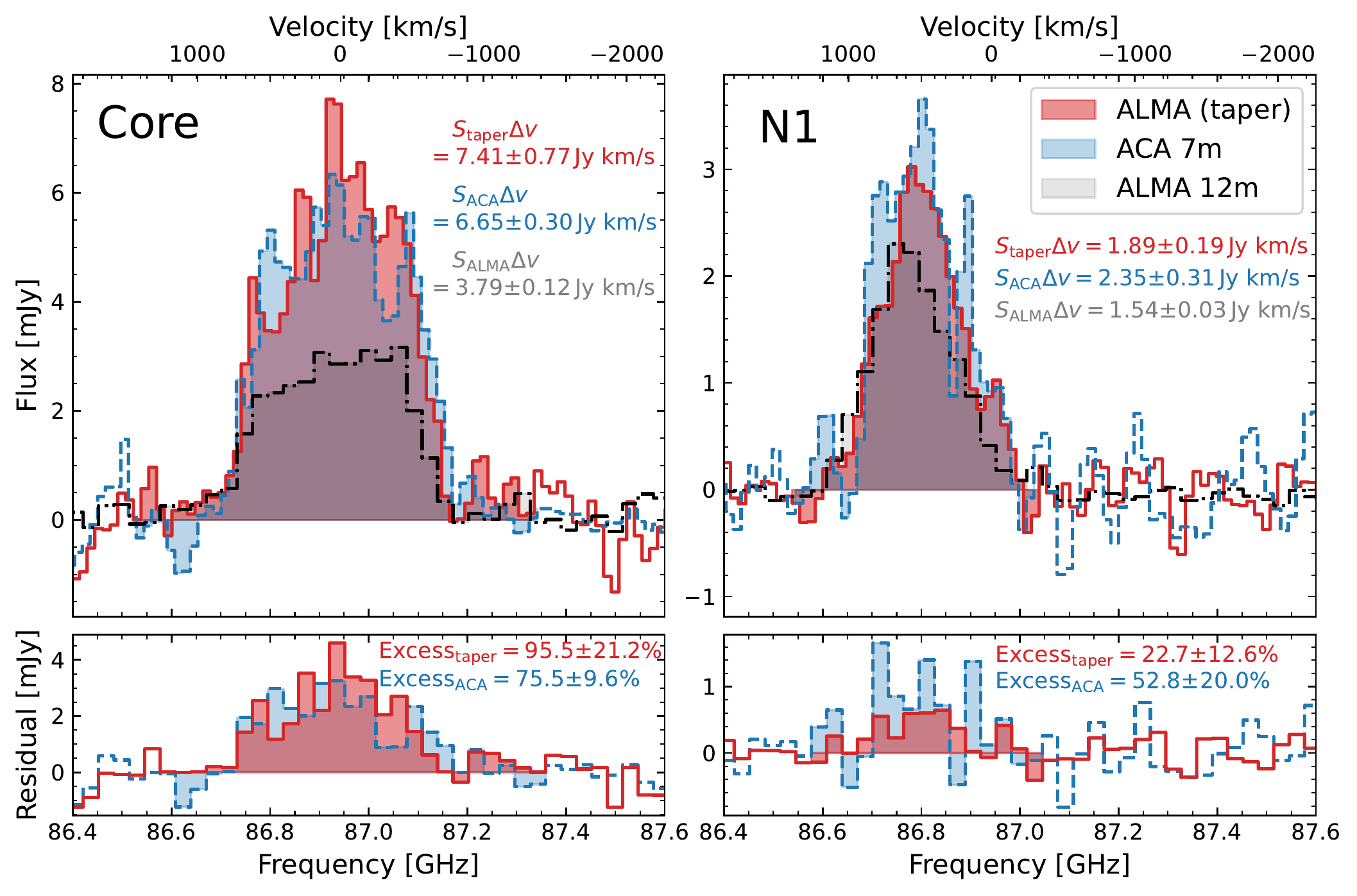}
     \caption{CO(4--3) spectra of core and `N1' from the tapered ALMA (red) compared to untapered ALMA (grey) and ACA (blue) data (top panels). The extra fluxes from the low-resolution results are plotted in the bottom panels. CO(4--3) excess is also shown in tapered ALMA data cube. }
     \label{fig:spec_taper}
\end{figure*}

We note that the $uv$-distance of the MRS ($d_{uv}\,{\sim}\,13.75\,\rm k\lambda$) of the ALMA observations is longer than its shortest baseline. 
To investigate if the extra diffuse emission can still be captured by ALMA, we tapered the ALMA data to a lower-resolution by 10\arcsec\ (Fig.\,\ref{fig:taper} right) and obtained the spectra of the southern core and `N1' (Fig.\,\ref{fig:spec_taper}). 
We measured their corresponding integrated CO(4--3) fluxes and continuum flux densities in an identical manner as in Section\,\ref{subsec: aca} (Table\,\ref{tab:taper}). 
Despite a larger uncertainty, the values of the CO(4--3) fluxes and 3.2\,mm dust continuum flux densities are consistent with the ACA results. 
We concluded that the extended gas reservoir can still be recovered by current available ALMA data, which requires an aggressive $uv$-tapering. 
\begin{deluxetable}{ccccc}[htp!]
\tablewidth{0pt} 
\tablenum{3}
\tablecaption{CO(4--3) fluxes and 3.2\,mm flux densities of the southern core and `N1' source measured from high-resolution ALMA, tapered ALMA, and ACA data. \label{tab:taper}}
\tablehead{
    Image & \colhead{$S_{\rm CO(4-3),core}\Delta v$} & \colhead{$S_{\rm 3.2mm,core}^{*}$} & \colhead{$S_{\rm CO(4-3),N1}\Delta v$} & \colhead{$S_{\rm 3.2mm,N1}^{*}$}\\  
    & \colhead{[Jy\,km/s]} & \colhead{[$\mu$Jy]} & \colhead{[Jy\,km/s]} & \colhead{[$\mu$Jy]}\\
    (1) & (2) & (3) & (4) & (5)}
\startdata 
    ALMA & 3.79$\pm$0.12 & 754$\pm$30 & 1.54$\pm$0.03 & 201$\pm$11 \\
    ALMA (tapered) & 7.41$\pm$0.77 & 667$\pm$48 & 1.89$\pm$0.19 & 211$\pm$55 \\
    Tapered excess [\%] & 95.5$\pm$21.2 & $-11.5\pm7.1$ & 22.7$\pm$12.6 & 4.8$\pm$28.0 \\
    ACA & 6.65$\pm$0.30 & 654$\pm$29 & 2.35$\pm$0.31 & 310$\pm$38\\
    ACA excess [\%] & 75.5$\pm$9.6 & $-13.2\pm5.1$ & 52.8$\pm$20.0 & 54.2$\pm$20.7 \\
\enddata
        \tablecomments{(1): Image used for the measurements. (2): Integrated CO(4–3) line strength of the southern core within $\pm3\sigma$ of the central velocity. The reported uncertainty is the primary beam-corrected rms value. (3): 3.2\,mm continuum flux density of the southern core within a kron aperture obtained by {\tt SourceExtractor}. The uncertainty is obtained from the primary beam-corrected standard deviation of values from 1000 random kron aperture. (4): Integrated CO(4–3) line strength of `N1'. (5): 3.2\,mm continuum flux density of `N1' within a kron aperture. \\$^{*}$: The source displays a complex morphology. We chose the aperture flux as the result. }
\end{deluxetable}
\bibliography{sample631}{}
\bibliographystyle{aasjournal}
\end{document}